\documentclass[11pt,twoside]{article}

\usepackage{asp2006}
\usepackage{epsf}
\usepackage{lscape}
\usepackage{graphicx}

\begin{document}
\title{Disk-Halo Interplay in Galaxy Evolution}
\author{Isaac Shlosman}    
\affil{Department of Physics \& Astronomy, University of Kentucky,
Lexington, KY 40506, USA; email: shlosman@pa.uky.edu}    

\begin{abstract}  
Some aspects of disk-halo interactions for models of in and out of equilibrium disk galaxies
are reviewed. Specifically, we focus on disk-halo resonant interaction without and in the presence
of a gas component. Another issue is the disk growth within an assembling triaxial dark matter 
halo. We argue that while the triaxiality is the result of the merger process and the radial
orbit instability, it is the developing chaos that damps the first generation of bars and
washes out the halo prolateness. This chaos is triggered by the gravitational quadrupole
interaction(s) in the system and supported by a number of other processes which are characteristic
of baryons.
\end{abstract}

\section{Introduction}   

The current paradigm of galaxy formation necessitates that galactic disks grow and evolve
within dark matter (DM) halos. As such, the baryonic disks serve as a test bed
for studying the DM properties, its dynamics and morphology, as well as mass,
momentum and energy exchange within the disk-halo system. Interaction between the galactic
disks and DM halos can in principle involve an exchange of these quantities.  
For example, observations of the low angular momentum H\,I gas, deep in the halo of NGC~891, 
require 
cold extragalactic gas influx (e.g., Fraternali et al. 2007), maybe through filaments
(Dekel \& Birnboim 2006). On the
other hand, the X-ray halos of starbursts and some normal galaxies can be explained
by the supernovae-heated gas driven from the disk or by the shock-compressed gas in the 
halo (e.g., Strickland et al. 2004). 

The situation is much more straightforward with the angular momentum ($J$) flow in the 
system. Galactic disks are rotationally supported, while the DM halos have low $J$, and, 
therefore, a low spin parameter $\lambda$ (e.g., Barnes \& Efstathiou 1987; Frenk, White
\& Efstathiou 1988). Most of the disk galaxies are barred, 
especially in the NIR (e.g., Knapen, Shlosman \& Peletier 2000), with bar properties which 
remain steady
up to the redshift of $\sim 1$ at least (Jogee et al. 2004; Elmegreen, Elmegreen \& Hirst
2004), and 
except for the very early Hubble types, they exhibit a spiral structure. This prevailing
{\it disk asymmetry} is extremely important for enhancing the $J$ transfer
between different morphological components --- the alternative way can be achieved, e.g., by 
dynamical friction of baryons against the DM. The current understanding of galactic bar
formation relies on the spontaneous breakup of the axial symmetry in the disk, the so-called
classical bar instability (e.g., Hohl 1971), originally applied to isolated disks.
Paradoxically, while Lynden-Bell \& Kalnajs (1972) have shown that bars and
spirals facilitate the $J$ transfer, DM halos have been considered
as the main stabilizers against the bar instability for years (Ostriker \& Peebles 1973), 
though they appear rather to be sinks of the disk momentum (e.g., Athanassoula \& 
Misiriotis 2002).

Within the framework of the classical bar instability, a stellar bar can develop only if
the unstable region loses its $J$. Gravitational torques serve as the mechanism that 
drives this
process (Lynden-Bell \& Pringle 1974). Their action can be described in terms of a non-local
viscosity (Lin \& Pringle 1987; Shlosman 1991) which shortens the characteristic timescale 
of the $J$
transfer dramatically, making this a dynamical rather then secular process. In principle,
$J$ can flow across the corotation radius (CR) to the outer disk, to the DM halo, or to
a flyby galaxy in case of a tidal galaxy interaction. The first option limits the $J$
flow because the mass of the outer disk is typically $\sim 20\%$ of the disk mass and its
ability to absorb $J$ quickly saturates (Fig.~1, left). On the other hand, the halo is 
massive, its inner part has a comparable mass to the inner disk, and its overall $J$ is 
small --- the halo is not supported by rotation. Fig.~1 shows also that the outer 
halo, beyond the disk radius, is fully susceptible to the $J$ transfer, after the inner 
halo efficiency has decreased. 

\begin{figure}[!t]
\plotfiddle{fig01a.ps}{3.5cm}{-90}{45}{45}{-170}{140}
\plotfiddle{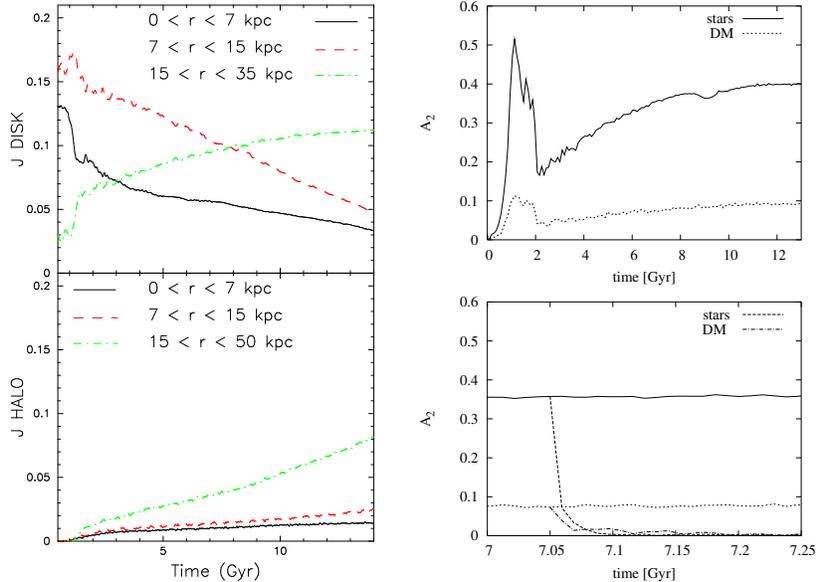}{3.5cm}{0}{31}{31}{-10}{-16.5}
\caption{
\underline{\it Left:} Evolution of angular momentum, $J$, in the disk (upper panel) and
the halo (lower panel). Note, $J$ saturates in the disk outside the bar CR (from
Martinez-Valpuesta, Shlosman \& Heller 2006). \underline{\it Right:} The $m=2$ amplitude,
$A_2$, of the stellar and DM ghost bars (upper panel) evolution. The fast dissolution of
the ghost bar (lower) after its stellar counterpart is axisymmetrized. Note the very
short time period for the lower frame (I. Berentzen \& I. Shlosman, unpublished)}
\label{fig1}
\end{figure}

The relatively local density response in the DM halo to the growing bar in the disk is 
to generate
a shadow `ghost' bar in the DM (Athanassoula 2006, 2007; Berentzen \& Shlosman 2006).
The ghost has the pattern speed of the stellar bar but its mass distribution differs
and is more centrally concentrated. The particle orbits in the DM ghost, therefore, are
in resonance with the stellar orbits. The principal difference between these two bars
is that the DM bar is not strictly speaking a bar at all --- it is not self-gravitating
{\it per se} and represents a gravitational wake induced in the DM by its stellar
counterpart. When the stellar bar is axisymmetrized, the DM bar dissolves in a fraction
of a crossing time (Fig.~1, right).  

\section{Angular momentum transfer in collisionless disk-halo systems}

The drain of $J$ from the inner disk appears to proceed selectively --- some of the orbits 
lose much
more than their neighbors (Athanassoula 2004; Martinez-Valpuesta, Shlosman \& Heller 2006). Fig.~2 
displays this phenomenon --- stellar/DM particles populating 
orbits with specific 
frequencies $\nu\equiv (\Omega-\Omega_{\rm b})/\kappa$ which correspond to lower resonances,
0, $\pm$ 1:2, 1:3, 1:4, etc. dominate the $J$ exchange. Here 
$\Omega$ is the angular velocity, $\Omega_{\rm b}$ is the bar pattern speed, and $\kappa$ is
the radial epicyclic frequency. More precisely, the particles which are more affected are 
those ultimately trapped by the above resonances. The dominant resonance in the disk
which emits $J$ is the inner Lindblad resonance (ILR), and the dominant resonance in the 
halo which absorbs it is the CR. So the direction of the $J$ transfer is confirmed to be 
from the inner disk to the halo. Fig.~10 of Martinez-Valpuesta et al.
(2006) displays a `forest' of lower resonances in the halo outside the CR which 
appear to be active in absorbing $J$ from the disk.

\begin{figure}[!ht]
\begin{center}
\includegraphics[angle=0,scale=0.65]{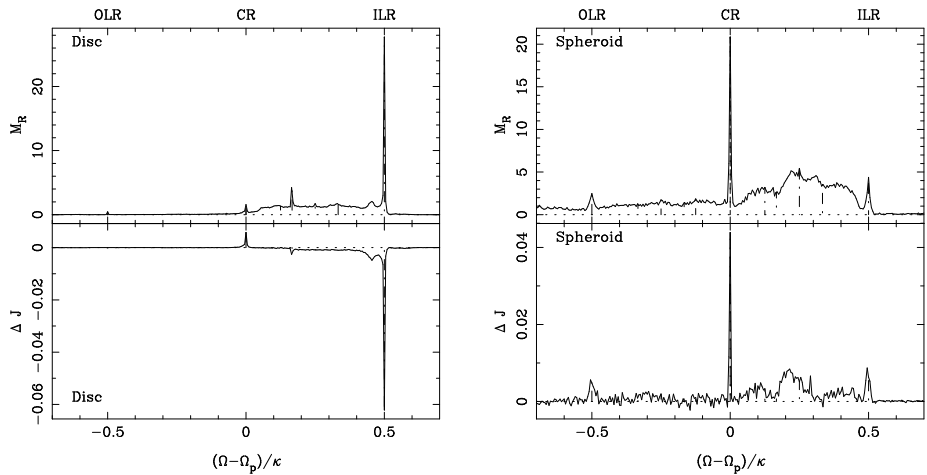}
\end{center}
\caption{Resonances and $J$ exchange. {\it Upper panels:} disk or halo mass per unit 
normalized frequency $\nu\equiv (\Omega-\Omega_{\rm b})/\kappa$ as a function of $\nu$. 
{\it Lower panels:} $\Delta J$ 
by the particles as a function of $\nu$. {\it Left panels:} disk, right panels: DM halo.
The vertical dash-dotted lines give the positions of the main resonances (ILR $\rightarrow$ 
0.5, CR $\rightarrow$ 0, OLR $\rightarrow$ -0.5, etc). From Athanassoula (2004). 
}\label{fig2}
\end{figure}

The bar growth is closely related to the $J$ loss from the disk within the CR. All
orbits within the bar are near-resonant, i.e., trapped by the resonances, mostly by the
ILR. As the resonances sweep across the phase and configuration spaces, because the bar
slows down, disk particles can lose their $J$ abruptly, being added (mostly) to the outer bar
in this process (Martinez-Valpuesta 2006). Most of the untrapped particles by the bar
can be found between its end and the CR. Bars that do not slow down, also do not grow
in length. For the trapped particles, much of the $J$ evolution has ceased.

While Athanassoula (2004) and Martinez-Valpuesta et al. (2006) have frozen the
gravitational potential in order to integrate the particle orbits, Ceverino \& Klypin (2007)
used live potential for the relatively short time period of 1~Gyr. It is
not clear whether the latter method provides any advantage because the short integration
time results in broader and less defined resonances. 

Different halos are expected to exhibit various efficiencies in their resonance interaction
with the disk. So far, we are aware of two factors which govern this process --- the
velocity dispersion in the halo and its mass density near the lower resonances (Athanassoula
2003), however, additional dependencies are expected. 

Cosmological simulations of DM structure show that the halos tend to acquire a universal
density profile, $\rho(r)$,  which is characterized by a central cusp with $\rho(r) \sim r^{-1}$ 
(e.g., Navarro, Frenk \& White 1997, hereafter NFW). This trend is challenged by
observations of galactic rotation curves that typically indicate the presence of a central 
flattening, i.e.,
core, instead of the cusp (e.g., Bosma 2004), and by a theoretical estimate of dynamical friction 
of a stellar bar against the DM background, which requires a density core to explain why
bars are still spinning (e.g., Sellwood 2006, 2007). But what is the possible cause of the DM core?

A number of alternatives have been explored for this purpose. Weinberg \& Katz (2007)
have advanced the idea that stellar bars transfer $J$ to the DM cusp and flatten it.
However, the usage of an analytical bar and other conditions which prevent the system to
respond in a self-consistent way, make their arguments unconvincing, especially the need
for a large number of particles, $N > 10^8$, to resolve the bar-cusp interaction (e.g.,
Sellwood 2007). Live bars do not dissolve the DM cusps in the axisymmetric halos even with 
effectively $N\sim 10^{10}$ (Dubinski,
Shlosman \& Berentzen, in preparation), but see Holley-Bockelman, Weinberg \& Katz (2005).
A  possibility that the supernova feedback injected energy flattens the
cusp has been proposed (Mashchenko, Couchman \& Wadsley 2006).
The other option is that the DM cusp has
been flattened prior or during the disk formation, when massive baryonic clumps descended
to the center via dynamical friction (El-Zant, Shlosman \& Hoffman 2001; Tonini, Lapi \&
Salucci 2006). It should be
pointed out that the cusp existence requires a `temperature' inversion, when solving
the Jeans equations without the central velocity anisotropy. Such an anisotropy was 
analyzed by Dehnen \& McLaughlin (2005). In the absence of the DM velocity anisotropy, 
the cusp is thermodynamically unstable, and only the lack of an 
energy flow in the collisionless matter allows for its existence --- a condition which
is resolved by the dynamical friction. Similarly, the DM cusps in galaxy clusters can be
erased by massive baryonic clumps --- those are played out by individual galaxies (El-Zant
et al. 2004).  

\section{Angular momentum transfer in the presence of gas}

While the $J$ redistribution in the collisionless disk-halo system has many additional caveats,
it is interesting to explore the effects of an additional component --- the gas. Currently,
the gas fraction in disk galaxies probably does not exceed 10\%, but the gas is known
to influence the galaxy evolution well beyond its fractional mass. Moreover, the gas fraction
in disks at redshifts above unity can be substantially higher, left alone without the
dissipative effects in the gas the disk would not form in the first place. 

\begin{figure}[!t]
\plotfiddle{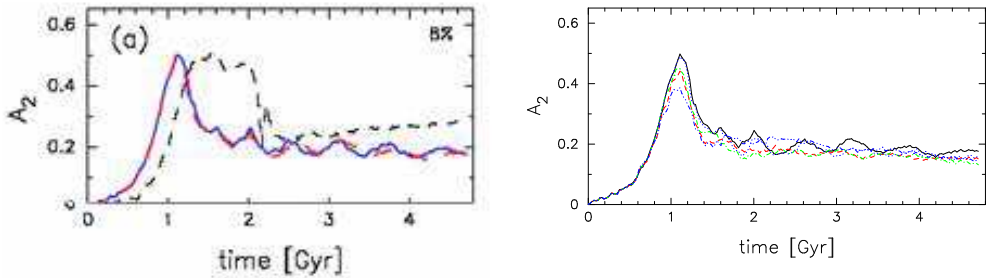}{1.7cm}{0}{106}{106}{-245}{-393}
\plotfiddle{fig03b.ps}{1.7cm}{-90}{33}{33}{-20}{123}
\caption{\underline{\it Left:} Evolution of the stellar bar ($m=2$ mode amplitude $A_2$) in 
models with $f_{\rm g}=0\%$ (solid line) and 8\% (dashed) over the first 5~Gyr.
\underline{\it Right:} Evolution of $f_{\rm g}=8\%$ model (solid lane) and the test models with 
gradually subtracted gravitational torques from the gas onto the stars --- 25\% subtracted 
(dotted), 50\% (dash-dotted), 75\% (dashed) and 100\% (dash-dot-dot-dotted). From Berentzen 
et al. (2007).
}\label{fig3}
\end{figure}

Fig.~3 (left panel) displays the disk-halo evolution in the presence of various gas fractions,
$f_{\rm g}\sim 0\%-8\%$ (Berentzen et al. 2007). Similar bars develop in all models and experience a
vertical buckling that `heats' up the stellar `fluid' along the rotation axis, weakening
the bar. Although $f_{\rm g}$ differs dramatically among these models, the  
evolution of the bar strength, $A_2$, shows the same 
maximum and nearly identical decrease in $A_2$. This decrease has been mistakenly
attributed to the $J$ transfer from the gas to the stars in the
bar by Bournaud, Combes \& Semelin (2005). On the other hand, the extent of the plateau
in $A_2$ depends strongly on $f_{\rm g}$ and clearly anti-correlates with it.

\begin{figure}[!t]
\plotfiddle{fig04a.ps}{1.7cm}{-90}{32}{26}{-200}{81}
\plotfiddle{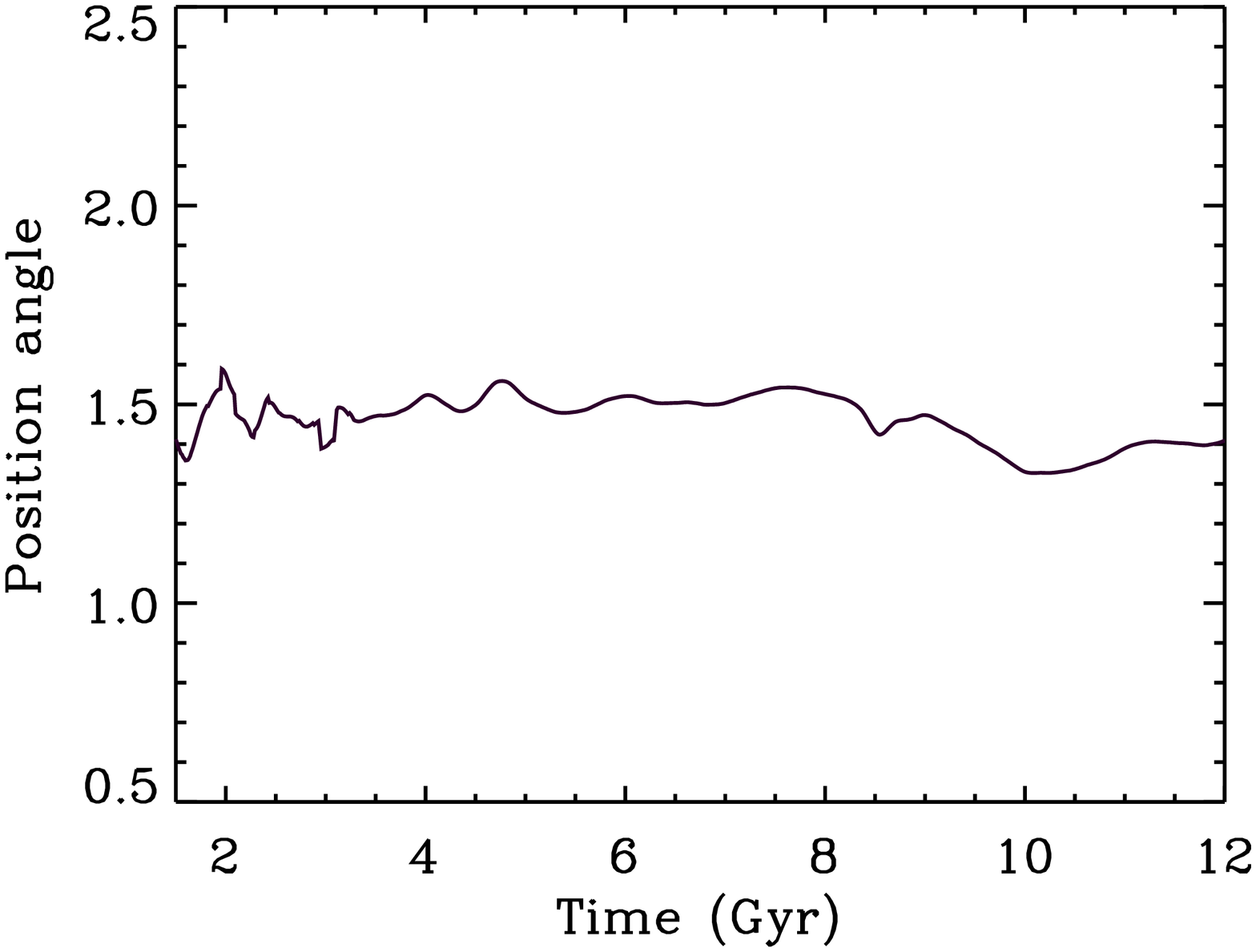}{1.7cm}{0}{28}{23}{15}{-46}
\caption{\underline{\it Left:} Change in the angular momentum, $\Delta J$, within 
the bar CR of the 8\% gas model shown in the left frame of Fig.~3. The solid line shows
$J$ lost by the stars, calculated directly from the model evolution. The
thin dashed line shows $J$ acquired by the stars from the gas, calculated from
integration of the corresponding gravitational torques in the simulation (it was shifted 
down in $y$ to match the solid line), and vice versa, from the stars to the gas (dotted 
line). The thick dashed
line provides the $J$ flow in the gas component, obtained from the model evolution
(from Berentzen et al. 2007). \underline{\it Right:} Position angle (in radians) of the
major axis of the DM halo in a cosmological simulation of a disk/halo formation in the 
$\Lambda$CDM Universe with the WMAP3 parameters (from Romano-Diaz et al., in prep.).
}\label{fig3}
\end{figure}

To show explicitly that the $J$ transfer from the gas does not alter the bar evolution
over the first 5~Gyr, Berentzen et al. (2007) have run a number of test models,
specifically removing the gravitational torques from the gas onto the stars and DM
(Fig.~3, right panel). The resulting $A_2$ curves are very similar. This test has been
supplemented by the direct analysis of $J$ redistribution in the disk halo system.
The inner (inside the CR) disk loses its $J$ to the outer disk and the DM halo. 
The inner halo absorbs the disk $J$ and forms the ghost bar. 
Fig.~4 (left frame) shows that the gas inputs into the stars $\sim 13\%$ of $J$ lost 
by the disk stars to the halo. The overall $J$ balance in the system is hardly affected
by the gas presence.
Surprisingly, {\it the halo is capable of absorbing $J$ from the disk on a relatively
short timescale}. This timescale is defined by the gas, while the bar evolution
is not. 
Bournaud et al. (2005) concluded that the input of $J$ by the gas torques contributes
to the bar dissolution --- the halo used in this work was frozen and, therefore, not
allowed to participate in the overall $J$ exchange. Note that the sharp decrease in
$A_2$ of the bar is not related to the angular momentum at all --- it results from
the buckling instability.

\section{DM halo shapes and disk morphology: the source of chaotic dynamics}

The angular momentum redistribution in the disk-halo system is facilitated by the
departure from axial symmetry. It is of paramount importance, therefore, that 
cosmological simulations universally lead to the formation of triaxial DM halos, i.e.,
both prolate and flattened (e.g., Bullock 2002; Allgood et al. 2006). Hence we
can expect high-$z$ halos to follow this trend. On the other hand, observations
indicate that halos in the nearby Universe are rather oblate and somewhat flattened,
based on a number of arguments, e.g., the coherence of the Sagittarius tidal stream 
(Ibata et al. 2001), polar rings (Sparke 1986; although their shape can be affected
by their self-gravity), diffuse X-ray emission around ellipticals (Buote et al. 2002),
H\,I isophotes of galactic disks (Merrifield 2002), weak lensing (Hoekstra, Yee \& Gladders 
2004), etc. The main issue, therefore, is {\it how the prolate DM halos at high redshifts
have been transformed into oblate ones in the local galaxies}. Both mass, angular 
momentum and energy exchange between the disk and the halo are expected to contribute 
to this process. 

While triaxial collisionless systems are not exactly in equilibrium and are expected 
to evolve secularly, the
addition of baryons accelerates this process dramatically and reduces the asymmetry
(Dubinski 1994; Kazantzidis et al. 2004). However, these works did not 
address the reason(s) for the shape evolution. 

Why do baryons reduce the halo triaxiality so dramatically and how is this effect related
to the internal dynamics of the system? Even in
pure DM halos, within the hierarchical merging framework, rich substructure forms
in the $\Lambda$CDM cosmology. These DM clumps join the main halo mostly along filaments.
The `thermalization' of such streamers and the accompanied dynamical friction that the clumps
experience when deep inside the main halo are themselves
irreversible processes that randomize the DM trajectories. The clumps also transfer $J$ to
the smooth DM component --- again changing the DM orbit shapes (El-Zant et al. 2001, 2004;
Tonini et al. 2006). The resulting inhomogeneities
inside the halo serve as multiple isotropic scattering centers. This process, while being
inherently related to the disk formation, in fact, precedes it. In elliptical galaxies it
may dominate the reshaping of the DM figure. Baryons only enhance this
trend. Dissipation leads to their accumulation in the inner halo, thus increasing the
baryon-to-DM mass ratio there and making them more dynamically important. If the baryons
are allowed to collapse to a disk, they become rotationally supported. But within
the prolate background potential, the disk acquires a substantial quadrupole moment,
as axisymmetric trajectories do not exist under these conditions.

A new and interesting phenomenon occurs because the halo figure tumbles slowly while
the baryons in the disk are rotationally supported (Shlosman 2007; Heller, Shlosman \& 
Athanassoula 2007b; Romano-Diaz et al., in prep.). Under these conditions, the inner
ILR (if it exists) approaches the center, while the outer ILR moves to large radii outside
the halo. This assures that the baryons in the disk respond out of phase with the DM
potential --- the required response of trajectories between the ILRs.
Support for the halo potential shape weakens and its prolateness is diluted.
{\it The near absence of halo figure rotation has important consequences
for the dynamics of galactic disks.}

\begin{figure}[!t]
\begin{center}
\includegraphics[angle=0,scale=0.5]{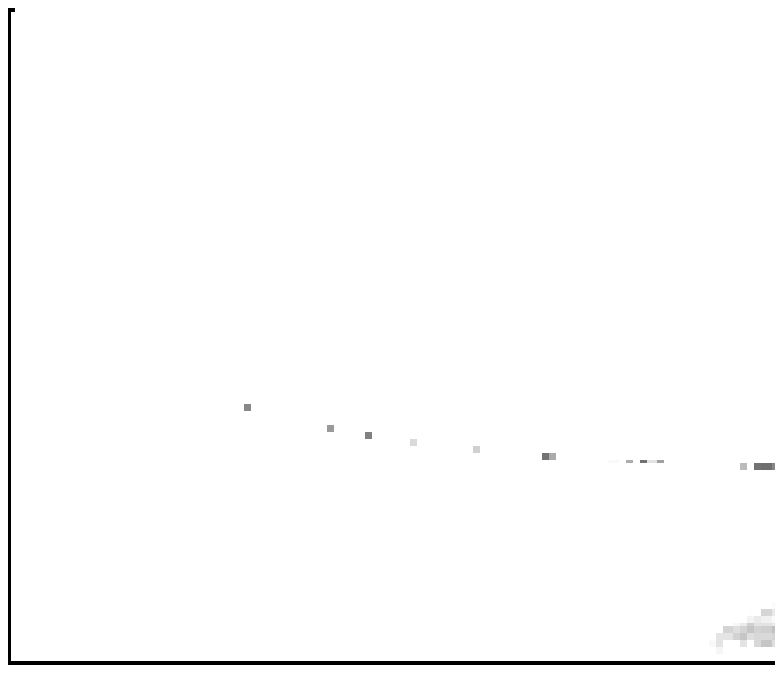}
\includegraphics[angle=0,scale=0.503]{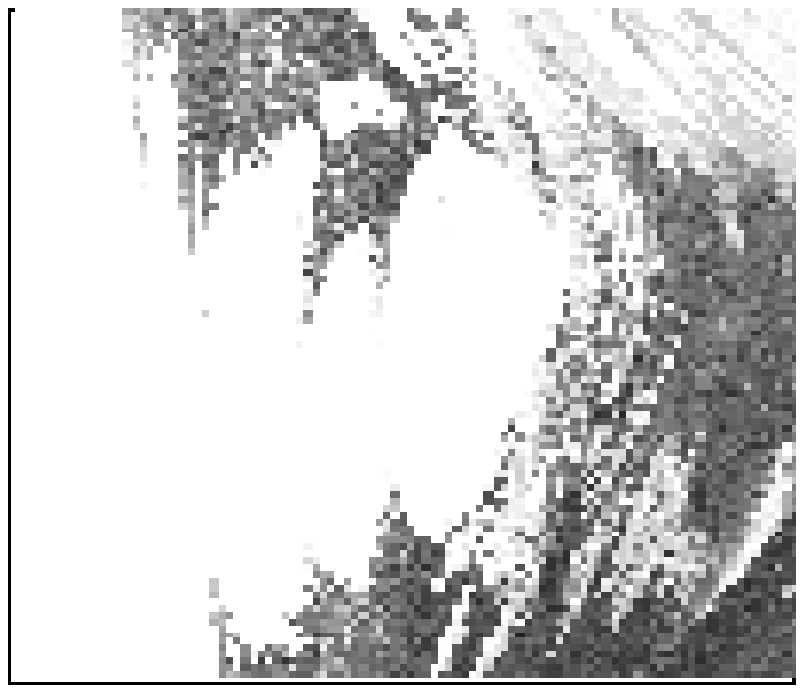}
\includegraphics[angle=0,scale=0.53]{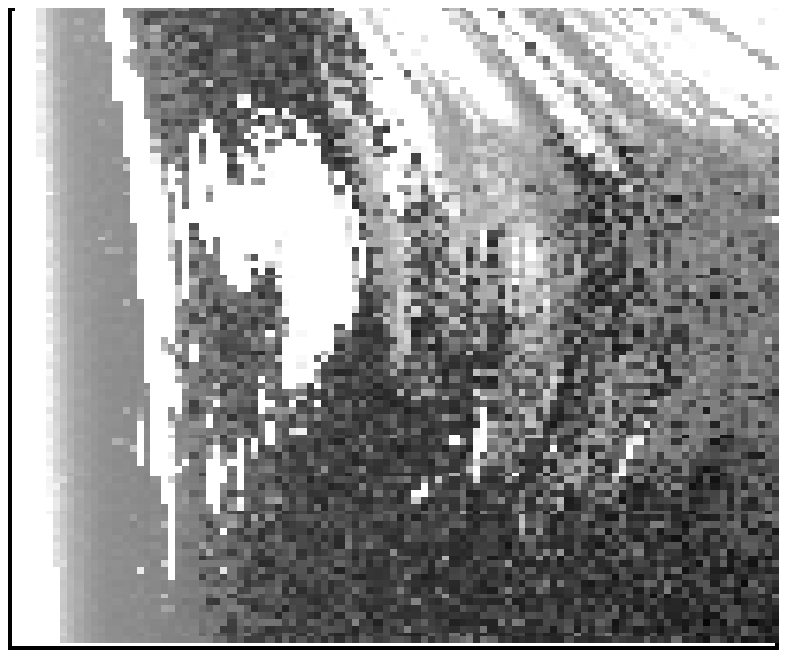}
\end{center}
\caption{Comparison of developing chaos in various systems given by gray-scale
diagrams showing the maximal Liapunov exponents at 50~Gyr. The x-axis gives the
cylindrical distance from the rotation axis and extends to 6~kpc. The y-axis gives
the particle velocities normalized by the local circular velocity, $v/v_{\rm c}$, and
extends to $v/v_{\rm c}=1.2$. The frames display
the Liapunov exponents for the particle
trajectories in the system of \underline{\it Left:} a pure triaxial DM halo with a
core radius of 5~kpc and an equatorial potential ellipticity of 0.1.
\underline{\it Middle:} an axisymmetric halo with a core radius of 5~kpc, and with
a stellar bar of $\sim 5$~kpc. \underline{\it Right:} triaxial halo with a core radius
of 5~kpc and the equatorial potential ellipticity of 0.1, and with a 5~kpc stellar bar.
The gray-scale is logarithmic with the white background corresponding to a characteristic
time of $10^6$~yr or less, and the darkest background --- to $10^{10}$~yr or more
(from El-Zant \& Shlosman 2002).
}\label{fig5}
\end{figure}

Even more important is the specifics of the baryon response to the halo prolateness ---
large-scale shocks induced in the growing and gas-rich disk by the gravitational torques
--- which act as a finite perturbation from the halo onto the disk. As a result, the
disk will be strongly asymmetric as long as the halo maintains its equatorial
prolateness within the disk region. The halo and disk quadrupoles will acquire different
pattern speeds (Heller, Shlosman \& Athanassoula 2007a,b). Their mutual interaction
will lead to chaos and dissolution of the least massive quadrupole, whether this is
a spontaneous or induced bar (El-Zant \& Shlosman 2002). The latter work found that the 
presence of a large-scale
stellar bar in the disk immersed in a mildly prolate halo triggers a chaotic behavior in the least
massive component --- the bar. Using the Liapunov exponents as a measure of chaos in the
bar, El-Zant \& Shlosman found that the stellar trajectories quickly become chaotic, dissolving the
bar. When the slowly tumbling nonaxisymmetric perturbation in the {\it potential} has
reached 10\%, almost all trajectories integrated in time were found to be chaotic and have large
Liapunov exponents. More centrally concentrated halos appear to be more dominated by chaos.

Fig.~5 quantifies the developing chaos in terms of the Liapunov exponents in three specific 
cases of 5~kpc flat density core DM halos. Interestingly, a pure DM triaxial halo, with no
bar, is nearly 
regular (Fig.~5, left panel), and no chaos is associated with this potential over the time period 
exceeding the Hubble time. An axisymmetric halo which hosts a bar exhibits a healthy amount
of chaos around the stellar (and ghost) bar CR, as expected (Fig.~5, middle panel). It is the 
addition of the
fast tumbling bar within a prolate halo that generates {\it continuous} chaos within the CR
and at most of the velocities (Fig.~5, right panel). The
characteristic timescale for the trajectories divergence is much less than the Hubble time.
The halo core potential can then be approximated as a quadratic form where the motion is separable
in Cartesian coordinates. Centrally concentrated
potentials do not have such symmetries --- the oscillations in the different degrees of freedom
are, in general, coupled, and no global integrals of motion exist. This makes plausible a
situation where most trajectories in the central regions conserve only energy, in which case
self-consistent solutions become impossible.

Galaxies that are not
initially centrally concentrated may acquire a central mass concentration by accreting gas
via spontaneous or tidally-induced bars. The resulting coupling between the degrees
of freedom will destroy the integrals of motion for a large enough fraction of orbits and
dissolve the bar, which is replaced by e.g., a central mass concentration (CMC). This scenario 
requires massive CMCs,
e.g.., supermassive black holes (SBHs) in excess of $10^9~{\rm M_\odot}$, which are not confirmed by observations
(El-Zant et al. 2003).

Heller et al. (2007b) have compared the {\it shape} evolution
of an assembling DM halo with and without the baryons. Fig.~6 (left panel) shows the time evolution
of the pure DM shape from 2~kpc to 300~kpc. Fig.~6 (right panel) exhibits
the same evolution when the disk grows by converting the gas to stars during the collapse
of an isolated density perturbation embedded in the Hubble flow.
The halo triaxiality is reasonably stable with
time for the pure DM models at all radii --- only a mild secular decrease was found
over the Hubble time. Models with baryons show a different behavior within
and outside the disk radius (roughly 15~kpc). The inner halo prolateness is washed out
by the time a disk has formed, $\sim 2-3$~Gyr.
The outer halo prolateness and flatness go through an adjustment phase and then
stabilize at some value, remaining substantial.

\begin{figure}[!t]
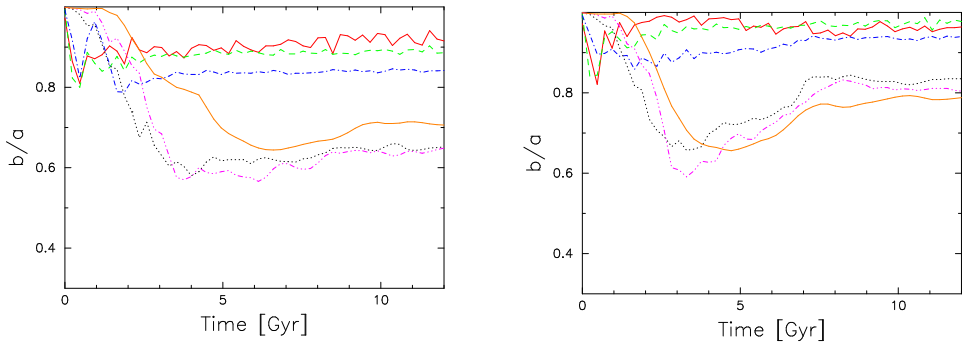

\plotfiddle{fig06a.ps}{1.9cm}{-90}{27}{27}{-215}{85}
\plotfiddle{fig06b.ps}{1.9cm}{-90}{27}{27}{-20}{150}
\caption{Evolution of DM halo intermediate-to-major axial ratio
$b/a$ for a pure DM model (left panel) and a DM with baryons model (right panel). The curves
correspond to 2~kpc (thick solid line), 5~kpc (dashed), 10~kpc (dot-dashed), 50~kpc
(thin solid), 100~kpc (dotted) and 300~kpc (dash-dot-dot-dotted). From Heller et al.
(2007b).
}\label{fig6}
\end{figure}

Berentzen, Shlosman \& Jogee (2006) have verified in live $N$-body simulations that the interaction of 
two or more quadrupoles leads to chaotic dynamics both in the disk and in the inner halo.
A typical decay time of stellar bars and disk ovalness was found
to lie in the range of a few Gyr, with a caveat that the halo itself suffers a reduction of its
quadrupole in the feedback, and, in extreme cases, the halo prolateness can be damped
ahead of that in the disk. More cuspy halos have been losing their prolateness within the radius 
faster, allowing the bar to survive.  This effect was further quantified in simulations with 
a tenfold growing stellar
disk embedded in the halo which assembles under cosmological initial conditions
within a 4~Mpc computational box (Berentzen \& Shlosman 2006).
More massive disks have erased the halo prolateness faster and more efficiently ---
hence, maximal disks are particularly apt to axisymmetrizing the background halos.
Moreover, the loss of prolateness by the halo allowed the maximal disk to develop
a next generation bar via classical bar instability, albeit at much later times. On the other
hand, disks dominated by the DM at all radii will have a much smaller effect,
if at all, on the shapes of the inner halo.

The emerging conclusion from the above works underlines the importance of self-consistent 
models in understanding the
DM halo shape evolution. If the disk is not allowed to generate substantial
quadrupole(s), or, in simple terms, is prevented from responding to the halo prolateness,
the basic ingredients for its feedback onto the halo are virtually eliminated.
There are three basic ways the disk responds to the halo: (1) its orbits are out-of-phase
with the halo potential (the disk is elongated normal to the halo major axis), 
(2) halo-induced and spontaneous stellar bars develop and tumble fast, and
(3) extensive stellar arms in the outer disk are triggered by the halo shape.
Recent interesting work by Debattista et al. (2007) has reached a different conclusion, 
based on 
simulations with a {\it frozen} disk which, therefore, cannot respond in any of the
above fashions. Furthermore, the disk is grown adiabatically, thus only reversible
processes are induced in the halo, making its orbits rounder. In fact, there is no basic
difference between slowly growing a rigid disk or a Plummer sphere within a responsive
halo --- the latter experiment was performed by Dubinski (1994). 
It is natural that Debattista et al. find that the chaos plays no role in reshaping the 
halo, and that only the (DM) substructure has an irreversible effect on the
halo shape.

To summarize, {\it the origin of the halo triaxiality lies to a certain degree in the major
merger process and
in the radial orbit instability, while the reduction of the halo triaxiality is
governed by developing chaos and transfer of the angular momentum to the DM}.
These processes can be facilitated through
the basic mechanism --- the separation of baryons from the DM that is driven by dissipation.
Specifically, (1) the cooling baryons enhance the substructure in the overall mass
distribution during
the halo assembly. The resulting clumps deposit energy and $J$ in the background DM
via dynamical friction --- this includes minor mergers and quiescent accretion and
is an irreversible process. (2) The ability of the baryons
to cool allows them to become more centrally concentrated (in the global and not only in
the local sense) --- the centrifugal barrier delays this process, by converting it
from a dynamical to a secular one. However, even the presence of this barrier does not assure
the dynamical-to-secular transition because the baryonic disk forming in the center will not
be axisymmetric in the presence of a prolate DM potential, and even if it will form
axisymmetrically, the classical bar instability will spontaneously break this symmetry.
(3) The resulting interaction of two or more quadrupoles which produce incommensurable
perturbing forces both on baryons and the DM will generate chaos in the disk and the
triaxial halo and act to reduce the triaxiality. (4) The central cusp, whether baryonic
or DM, can induce chaos as well, although this is not assured in the most general
sense.

\section{Evolving Disk Morphology: Bars, Bulges and All That}

The origin of galactic disks has emerged as one of the sticking issues in the overall
understanding of the galaxy formation process. Large gaps permeate our knowledge
of the origin of disk structural components, e.g., bars, bulges, SBHs, etc. The path to 
the observed tight correlation between
the properties of the central SBH and the surrounding bulge is unknown as well,
although suggestions abound. Some of the components require a certain degree
of dissipation to accompany their formation, such as the disk itself, certain
types of bulges, and the SBH --- this erases the memory of the initial conditions
and the link to cosmology. In particular, the origin timescale for current disk morphology
and the amount of evolution that occured since is only known remotely. A number of controversies 
have emerged: first, the angular momentum distribution of the observed (baryonic) disks 
does not 
match that of the disks obtained in cosmological numerical simulations --- the latter appear 
to have radial scalelengths which are too small (e.g., Sommer-Larsen, Gotz \& Portinari 
2002) --- the so-called angular 
momentum `catastrophe.' Moreover, the central bulges form too massive and there are 
difficulties to explain the widespread bulgeless disks. Second, the numerical simulations
of DM halo formation show that they display a universal density profile characterized
by a steep central density cusp (e.g., NFW), while 
observations point rather to a flat density core (e.g., de Blok \& Bosma 2002) ---
the core `catastrophe.' The central difficulty appears to be the conservation
of baryonic $J$ during the disk formation and major mergers events. The standard picture
of slow adiabatic collapse of a smooth baryonic component only aggravates the problem
by dragging the DM inward in an adiabatic compression.  

These and other discrepancies between observations and modeling cannot be fully 
resolved by the current theory --- additional high-resolution numerical simulations 
are required, focusing on the effect of various parameters of star formation, energy 
feedback from the stellar evolution and growing SBHs. Indeed, recent numerical 
modeling has 
made progress in understanding the complexity of disk formation and evolution.
Models with {\it frozen} spherically-symmetric DM halos, e.g., by Samland \& Gerhard 
(2003) and Immeli et al. (2004), have used a chemo-dynamical code to investigate the
bulge stellar population, as well as a possible bulge formation from massive
spiraling-in gas clumps, accompanied by a strong central starburst (see also Shlosman 
\& Noguchi 1993). Okamoto et al. (2005) have assumed that the star formation proceeds 
in two modes --- in a high density environment or shock-triggered, leading to
elliptical or disk galaxies, respectively. Hence, the star formation processes can
affect the final $J$ of the galaxy. Governato et al. (2004, 2007) have claimed to
produce disks with realistic exponential scalelengths, to fit the $I$-band and
baryonic Tully-Fisher relations, and to reduce the number of visible satellites
around the Milky Way-type halo, by combining the UV background and SN feedback.

Heller et al. (2007a,b) have focused on the formation of the disk and its structural 
components within the 
context of the collapse of an isolated density perturbation embedded in the Hubble flow.
In agreement with Fig.~4b, $J$ was channeled into the internal circulation,
leaving the DM halo figure tumbling $\sim \pi$ radian in a Hubble time.
At $z=0$, a range of nearly bulgeless-to-bulge dominated disks were obtained, with 
flat rotation curves which account for the observed disk/halo contributions.
A relatively large number of models was run to investigate the effect of star
formation and stellar evolution feedback onto the disk properties, normally unresolved
in cosmological simulations. Several single-varied
parameter evolutionary sequences have been constructed, such as varying
the energy thermalization parameter, which characterizes the feedback from stellar 
evolution, the star formation threshold parameter, the collapse time for the star forming
clouds, and the gravitational softening in the gas. A number of correlations
along these sequences have been found, e.g., the thermalization parameter appears to
correlate with the bulge prominence, and a lower density threshold leads
to smaller, thicker disks.  

While the bar origin lies in the $J$ loss by the inner disk, exponential bulges appear
to be largely by-products of the bar evolution. The vertical buckling instability
in a bar leads to the formation of boxy/peanut-shaped bulges (e.g., Combes \& Sanders 1981;
Combes et al. 1990). Although the bar weakens as a result (Raha et al. 1991; Pfenniger
\& Friedli 1991) and shortens, in principle it cannot be destroyed (Martinez-Valpuesta 
\& Shlosman 2004). Multiple bucklings of the bar add up to the growth of its bulge,
irrespective of the gas inflow (Martinez-Valpuesta et al. 2006). 

We have discussed earlier that the halo shape speeds up the $J$ exchange in the system.
This is especially pronounced in the early stages of disk growth, when its rotation
curve is probably dominated by the DM. The initial disk response to a finite perturbation
from the halo will naturally lead to a bar. The resulting mass cascade to the center,
e.g., in the form of nested bars (Shlosman, Frank \& Begelman 1989), can create the
environment for the formation and initial growth of a SBH (Begelman, Volonteri \&
Rees 2006). 
     
\section{Conclusions}

We have summarized some aspects of disk-halo interactions in equilibrium and forming
disk galaxies. Because the overall problem is of a high complexity, progress in 
this direction has been rather slow and dependent on numerical simulations. Nevertheless,
certain trends have emerged. The baryonic disk and DM halo exchange 
mass, angular momentum and energy during the initial formation phase as well as 
during the subsequent quiescent evolution. A number of processes, like star formation
as well as formation of the central SBH, require a much deeper understanding before
they can be modeled beyond the phenomenological approach. On the other hand, nonlinear
dynamics and the role of chaos in the evolution of disks and halos can be quantified
already at present.  

\acknowledgements 
I acknowledge illuminating discusssions with my colleagues, too numerous to list here. 
I am grateful to my collaborators Ingo Berentzen, Clayton Heller, Yehuda Hoffman, 
Inma Martinez-Valpuesta and Lia Athanassoula on these issues. The 
relevant grants from NASA and NSF are acknowledged. I also thank the 
conference organizers for generous support.

\end{document}